\documentclass[a4paper,twoside]{article}

\begin{document}

\title{On Some Accelerating Cosmological Kaluza-Klein Models}

\author{\bf Goran S. Djordjevi\'c and Ljubi\v sa Ne\v si\'c \\
\it Department of Physics, University of Nis, Serbia and Montenegro\\
E-mail: nesiclj, gorandj@junis.ni.ac.yu}

\date{}

\maketitle

\noindent {\bf Abstract. }We consider (4+D)-dimensional
Kaluza-Klein cosmological model with two scaling factors, as real,
$p$-adic and adelic quantum mechanical one. One of the scaling
factors corresponds to the D-dimensional internal space, and
second one to the 4-dimensional universe. In standard quantum
cosmology, i.e. over the field of real numbers $R$, it leads to
dynamical compactification of additional dimensions and to the
accelerating evolution of 4-dimensional universe. We construct
corresponding $p$-adic quantum model and explore existence of its
$p$-adic ground state. In addition, we explore evolution of this
model and a possibility for its adelic generalization. It is
necessary for the further investigation of space-time discreteness
at very short distances, i.e. in a very early universe.

\section{Introduction}

On the basis of the results of quantum gravity \cite{garray},
theoretical uncertainty of measuring distances is greater or equal
to the Planck distance. It leads to the conclusion that on very
short distances Archimedean axiom is not valid, i.e. the space can
posses ultrametric features. Because geometry is always connected
with a concrete number field, in the case of the nonarchimedean
geometry it is used to be the field of $p$-adic numbers $Q_p$. In
the high energy physics, these numbers have been used almost
twenty years. The motivation comes from string theory \cite{vol1}.
Generally speaking, $p$-adic approach should be useful in
describing a very early phase of the universe and processes around
Planck scale.

A significant number of papers, motivated by \cite{vol1}, has been
published up to now (for a review see \cite{vvz,brek}). In this
paper we are especially interested in application of $p$-adic
numbers and analysis in quantum cosmology \cite{vol2}. $p$-Adic
quantum mechanics \cite{mpla} (QM) has been successfully applied
in minisuperspace quantum cosmology \cite{grav}. We have treated
many cosmological models, mainly constructed in four space-time
dimensions \cite{vol2}. Based on that one can ask: is it possible
to extend this approach to the multidimensional quantum
cosmological models? In this article we briefly formulate such a
model.

After brief mathematical introduction in Section 2, a short review
of $p$-adic and adelic quantum mechanics and cosmology is given in
Section 3. Section 4 is devoted to the classical (4+D)-dimensional
cosmological models filed with "exotic" fluid \cite{dar}. The
corresponding $p$-adic model is considered in Section 5. This
paper is ended by short conclusion, including adelic
generalization, and suggestion for advanced research.

\section{$p$ -Adic Numbers and Adeles}

Perhaps the most easier way to understand $p$-adic numbers is if
one starts with the notion of norm. It is well known that any norm
must satisfy three conditions: nonnegativity, homogeneity and
triangle inequality. The completion of the field of rational
numbers $Q$ with respect to the absolute value, or standard norm
$|.|_\infty$, gives the field of real numbers $R\equiv Q_\infty$.
Besides this norm there are another ones which satisfy the first
two conditions and the third one in a stronger way
\begin{equation}
\label{norm} \|x+y\|\leq{\max}(\|x\|,\|y\|),
\end{equation}
so called strong triangle inequality. The most important of them
is $p$-adic norm $|.|_p$ \cite{vvz}. The feature (\ref{norm}),
also called ultrametricity, is one of the most important
characteristics of the $p$-adic norm ($p$ denotes a prime number).
The number fields obtained by completion of $Q$ with respect to
this norm are called $p$-adic number fields $Q_p$. It is known
that any nonzero rational number $x$ can be expressed as $x=\pm
p^\gamma a/b$, where a $\gamma$ is the rational number, and $a$
and $b$ are the natural numbers which are not divisible with the
prime number $p$ and have no common divisor. Then, $p$-adic norm
of $x$ is, by definition, $|x|_p=p^{-\gamma}$.

Because $Q_p$ is local compact commutative group the Haar measure
can be introduced, which enables integration. In particular, the
Gauss integral will be employed
\begin{equation}
\label{gauss} \int_{Q_p}\chi_p(\alpha x^2+\beta x)dx=
\lambda_p(\alpha)|2\alpha|_p^{-1/2} \chi_p \left(
-{\beta^2\over4\alpha} \right), \enskip\alpha\neq 0,
\end{equation}
\noindent where $\chi_p={\exp 2\pi i\{x\}_p}$, is an additive
character ($\{x\}_p$ is the fractional part of $x$), and
$\lambda_p(\alpha)$ is an arithmetic complex-valued function
\cite{vvz}.


It should be noted that there is the real counterpart of
$\lambda_p$
\begin{equation}
\lambda_\infty(\alpha) =
{1\over\sqrt2}(1-i\hbox{sign}\alpha),\quad \alpha\in Q_{\infty}.
\end{equation}
Commonly the main properties of $\lambda_v$, ($v$ denotes $\infty$
or any $p$) are:
\begin{equation}
\lambda_v(0) = 1, \quad \lambda_v(a^2\alpha) = \lambda_v(\alpha),
\end{equation}
\begin{equation}
\lambda_v(\alpha)\lambda_v(\beta) =
\lambda_v(\alpha+\beta)\lambda_v(\alpha^{-1}+\beta^{-1}), \quad
|\lambda_v(\alpha)|_\infty = 1.\label{2.3}
\end{equation}
Very simple but rather important function in $p$-adic analysis and
$p$-adic QM is
\begin{equation}
\label{2.8}
\Omega(|x|_p) =\left\{
\begin{array}{rl}
1,& |x|_p\le1,\\
0,& |x|_p>1,
\end{array}
\right.
\label{2.4}
\end{equation}
which is the characteristic function of $Z_p$. Note that
$Z_p=\{x\in Q_p: |x|_p\le1\}$ is the ring of $p$-adic integers.

Simultaneous treatment of real and $p$-adic numbers can be
realized by concept of adeles. An adele $a\in {\cal A}$ is an
infinite sequence $a=(a_\infty,a_2,\cdots,a_p,\cdots)$,
where $a_\infty\in R$ and $a_p\in Q_p$, with the
restriction that $a_p\in Z_p$ for all but a finite set $S$ of
primes $p$. The set of all adeles ${\cal A}$ can be written in the
form
\begin{equation}
{\cal A}=\mathop{U}\limits_{S}{\cal
A}(S), \quad {\cal A}(S) =
R\times \prod_{p\in
S} Q_p\times\prod_{p\not\in S} Z_p\
.\label{2.6}
\end{equation}
Also, ${\cal A}$ is a topological space. Algebraically, it is a
ring with respect to the componentwise addition and
multiplication. There is the natural generalization of analysis on
$R$ and $Q_p$ to analysis on ${\cal A}$ \cite{gel1}.

\section{$p$ -Adic and Adelic Quantum Mechanics and Cosmology}

$p$-Adic quantum mechanics is quantum mechanics over the field
$Q_p$, and $p$-adic quantum cosmology is the application of
$p$-adic QM on the universe as a whole \cite{vol2}.

\subsection{$p$ -Adic and Adelic Quantum Mechanics}

$p$-Adic QM has been developed in two different ways: in the first
one wave function is complex valued function of $p$-adic variable
\cite{vvz}, and in the second one, $p$-adic wave function depends
on $p$-adic variable \cite{khr}. If we want simultaneous (adelic)
treating of standard quantum mechanics and all $p$-adic's,
including usual probabilistic interpretation developed in standard
QM, then the first formulation is preferable one.

Because $p$-adic fields $Q_p$ are totally disconnected \cite{vvz},
dynamics of $p$-adic quantum models is described by a unitary
evolution operator $U(t)$ without using the Hamilton operator and
infinitesimal displacement. Therefore, $U(t)$ is formulated in
terms of its kernel ${\cal K}_t(x,y)$
\begin{equation}
\label{3.7} U_p(t)\Psi(x)=\int_{Q_p}{\cal K}_t(x,y)\Psi(y) dy.
\end{equation}
In the above equation ${\cal K}_t$ is the kernel of  the $p$-adic
evolution operator which is defined by the functional integral
(in units $G=\hbar=c=1$)
\begin{equation}
{\cal K}_t(x,y)=\int\chi_p \left( -\int_0^tL(\dot q,q)dt \right)
\prod_tdq(t),
\end{equation}
with properties analogous to the properties existing in standard
QM. In that way, $p$-adic QM is given by a triple \cite{vvz}
$(L_2(Q_p), W_p(z_p), U_p(t_p))$.
Keeping in mind that standard QM can be also presented as the
corresponding triple, ordinary and $p$-adic QM can be unified in
the form of adelic quantum mechanics \cite{dra1}
\begin{equation}
(L_2({\cal A}), W(z), U(t)),
\label{3.9}
\end{equation}
where $L_{2}({\cal A})$ is the Hilbert space on ${\cal A}$, $W(z)$
is the unitary representation of the Heisenberg-Weyl group on
$L_2({\cal A})$ and $U(t)$ is the unitary representation of the
evolution operator on $L_2({\cal A})$.

The adelic evolution operator $U(t)$ is defined as
\begin{equation}
U(t)\Psi(x)=\int_{{\cal A}} {\cal
K}_t(x,y)\Psi(y)dy=\prod\limits_{v}{} \int_{Q_{v}}{\cal
K}_{t}^{(v)}(x_{v},y_{v})\Psi^{(v)}(y_v) dy_{v}. \label{3.10}
\end{equation}
The eigenvalue problem for $U(t)$ reads
\begin{equation}
U(t)\Psi _{\alpha \beta} (x)=\chi (E_{\alpha} t) \Psi _{\alpha
\beta} (x), \label{3.11}
\end{equation}
where $\Psi_{\alpha \beta}$ are adelic eigenfunctions, $E_{\alpha
}=(E_{\infty}, E_{2},..., E_{p},...)$ denotes adelic energy,
indices $\alpha$ and $\beta$ denote  energy levels and their
degeneration, respectively. Any adelic eigenfunction has the form
\begin{equation}
\label{3.12}
\Psi_S(x) = \Psi_\infty(x_\infty)\prod_{p\in S}\Psi_p(x_p)
\prod_{p\not\in S}\Omega(| x_p|_p) , \quad x\in {\cal A},
\end{equation}
where $\Psi_{\infty}\in L_2(R)$ and $\Psi_{p}\in L_2(Q_p)$ are
ordinary and $p$-adic eigenfunctions, respectively. We would like
to paraphrase Manin's opinion \cite{manin} that our universe
cannot be described fully by real numbers nor $p$-adic's only,
than by adeles. In other words, our universe is inherently adelic.

\subsection{$p$ -Adic and Adelic Quantum Cosmology}

Adelic quantum cosmology can be understood as an application of
adelic quantum theory to the universe as a whole \cite{vol2}. In
the path integral approach to standard quantum cosmology, the
starting point is Feynman's path integral method, i.e. the
amplitude to go from one state with intrinsic metric $h_{ij}'$
and matter configuration $\phi'$ on an initial hypersurface
$\Sigma'$ to another state with metric $h_{ij}''$ and matter
configuration $\phi''$ on a final hypersurface $\Sigma''$ is
given by a functional integral
\begin{equation}
\langle h_{ij}'',\phi'',\Sigma''|
h_{ij}',\phi',\Sigma'\rangle_\infty =
\int {\cal D}{(g_{\mu\nu})}_\infty {\cal D}(\Phi)_\infty
\chi_\infty(-S_\infty[g_{\mu\nu},\Phi]),
\label {4.1}
\end{equation}
over all (four)-geometries $g_{\mu\nu}$ and matter configurations
$\Phi$, which interpolate between the initial and final
configurations \cite{wil1}. In above expression $S[g_{\mu\nu},\Phi]$ is
an Einstein-Hilbert action for the gravitational and matter
fields.

To perform $p$-adic and adelic generalization we first make
$p$-adic counterpart of the action using form-invariance under the
inter-change of the real to the $p$-adic number fields. Then we
generalize (\ref{4.1}) and introduce $p$-adic complex-valued
cosmological amplitude
\begin{equation}
\langle h_{ij}'',\phi'',\Sigma''|
h_{ij}',\phi',\Sigma'\rangle_p =
\int{\cal D}{(g_{\mu\nu})}_p{\cal D}(\Phi)_p
\chi_p(-S_p[g_{\mu\nu},\Phi]),
\label {4.3}
\end{equation}
where $g_{\mu\nu}(x)$ and $\Phi (x)$ are the corresponding
$p$-adic counterparts of metric  and matter fields continually
connecting
their values on $\Sigma'$ and $\Sigma''$.

The standard minisuperspace ground state wave function in the
Hartle-Hawking (no-boundary) proposal \cite{haw}, will be attained
if one performs a functional integration in the Euclidean version
of
\begin{equation}
    \Psi_\infty[h_{ij}] = \int {\cal D}{(g_{\mu\nu})}_\infty {\cal
D}(\Phi)_\infty
\chi_\infty(-S_\infty[g_{\mu\nu},\Phi]), \label {hh}
\end{equation}
over all compact four-geometries $g_{\mu\nu}$ which induce
$h_{ij}$ at the compact three-manifold. This three-manifold is the
only boundary of all the four-manifolds. If we generalize the
Hartle-Hawking proposal to the $p$-adic minisuperspace, then an
adelic Hartle-Hawking wave function is an infinite product
\begin{equation}
    \Psi[h_{ij}] =\prod_v \int {\cal D}{(g_{\mu\nu})}_v {\cal D}(\Phi)_v
\chi_v(-S_v[g_{\mu\nu},\Phi]), \label {adelichh}
\end{equation}
where the path integration must be performed over both,
archimedean and nonarchimedean, geometries. If evaluation of the
corresponding functional integrals gives as a result $\Psi[h_{ij}
]$ in the form (\ref{3.12}), we will say that such cosmological
model is adelic one.

A more successful $p$-adic generalization of the minisuperspace
cosmological models can be performed in the framework of $p$-adic
and adelic quantum mechanics without using the Hartle-Hawking
proposal. In such cases, we examine the conditions under which
some eigenstates of the evolution operator exist \cite{vol2}.

\subsection{Quantum Minisuperspace Models in $p$-Adic and Adelic Quantum
Mechanics}
\noindent

In the study of the homogeneous universes, the metric depends only
on the time parameter. It enables formulation of the models with a
finite dimensional configuration space, so called minisuperspace
\cite{wil1}. Their variables are the three-metric components, and
corresponding material fields, if any. The quantization of these
models can be performed adopting of the rules of QM. One of the
main reasons for using minisuperspace models in describing the
universe is their simplicity. While the full (superspace) quantum
cosmological models are usually unsolvable, the minisuperspace
ones can be handled with available mathematical tools.

In $p$-adic minisuperspace quantum cosmology, we investigate
necessary conditions for the existence of the $p$-adic ground
state in the form of $\Omega$-function. Some other typical
eigenfunctions like $\delta$-function can be also of interest. The
existence of these functions enables construction of $p$-adic
models and their adelization. Usually, this leads to the some
restrictions on the parameters for the many exactly soluble
minisuperspace cosmological models \cite{vol2}.

The necessary
condition for the existence of an adelic quantum model is the existence
of
$p$-adic  ground state $\Omega(|q_\alpha|_p)$  defined by (\ref{2.8}),
i.e.
\begin{equation}
\label{6.1}
\int_{|{q_\alpha}'|_p\leq1}{\cal K}_p
({q_\alpha}'',N;{q_\alpha}',0)d{q_\alpha}'=
\Omega(|{q_\alpha}''|_p),
\end{equation}
($q_\alpha$ are minisuperspace coordinates, $\alpha=1,2, ..., n$
and $N$ is the lapse function). Analogously, if a system  is in
the state $\Omega(p^\nu|q_\alpha|_p)$, where
$\Omega(p^\nu|q_\alpha|_p) = 1$ if $|q_\alpha|_p \leq p^{-\nu}$
and $\Omega(p^\nu|q_\alpha|_p) = 0$ if $|q_\alpha|_p > p^{-\nu}$ ,
then its kernel must satisfy equation
\begin{equation}
\label{6.2}
\int_{|{q_\alpha}'|_p\leq p^{-\nu}}{\cal K}_p
({q_\alpha}'',N;{q_\alpha}',0)d{q_\alpha}'=
\Omega(p^\nu|q_\alpha''|_p).
\end{equation}
If $p$-adic  ground state is of the form of the
$\delta$-function, where  $\delta$-function is defined as

$$
\label{delta} \delta (a-b) = \left\{
\begin{array}{rl}
1,& if\ \  a=b,\\
0,& if\ \  a\neq b,
\end{array}
\right.
$$
then  the corresponding kernel of the model has to satisfy
equation
\begin{equation}
\label{6.3} \int_{|q'_\alpha|_p =p^\nu}{\cal
K}_p(q_\alpha'',T;q',0)dq_\alpha'= \delta(p^\nu-|q_\alpha''|_p), \
\nu\in Z .
\end{equation}
Equations (\ref{6.1}) and (\ref{6.2}) are usual $p$-adic vacuum
ingredients of the adelic eigenvalue problem (\ref{3.11}), where
$\chi_p(Et)=1$ in the vacuum state $\{ Et\}_p = 0$. The above
$\Omega$ and $\delta$ functions do not make a complete set of
$p$-adic eigenfunctions, but they are very simple and
illustrative. Since these functions have finite supports, the
ranges of integration in (\ref{6.1})-(\ref{6.3}) are also finite.
The lapse function $N$ is under the kernel ${\cal
K}(q_\alpha'',N;q_\alpha',0)$ and $N$ is restricted to the some
values on which eigenfunctions do not depend explicitly.

\section{(4+D)-Dimensional Cosmological Models Over the Field of Real
Numbers}

The old idea that four dimensional universe, in which we exist, is
just our observation of physical multidimensional space-time
attracts much attention nowadays. In such models compactification
of extra dimensions play the key role and in the some of them
leads to the period  of accelerated expansion of the universe
\cite{dar, tow, jal}. This approach is supported and encouraged
with the recent results of the astronomical observations. We
briefly recapitulate some facts of the real multidimensional
cosmological models, necessary for $p$-adic and adelic
generalization. The metric of such Kaluza-Klein model with
D-dimensional internal space can be presented in the form
\cite{dar, wud}
\begin{equation}
ds^2=-\tilde
N^2(t)dt^2+R^2(t)\frac{dr^idr^i}{(1+\frac{kr^2}{4})^2}
+a^2(t)\frac{d\rho^a d\rho^a}{(1+k'\rho^2)^2}
\end{equation}
where $\tilde N(t)$ is a lapse function, $R(t)$ and $a(t)$ are the
scaling factors of 4-dimensional universe and internal space,
respectively; $r^2\equiv r^i r^i (i=1,2,3)$, $\rho^2\equiv\rho^a
\rho^a (a=1,... D)$, and $k,k'=0,\pm 1$. The form of the
energy-momentum tensor is
\begin{equation}
T_{AB}=diag(-\rho,p,p,p,p_D,p_D,...,p_D),
\end{equation}
where indices $A$ and $B$ run over both spacetime coordinates and
the internal space dimensions. If we want the matter is to be
confined to the four-dimensional universe, we set all $p_D=0$.

Now, we examine the case for which the pressure along all the
extra dimensions vanishes $p_D=0$ (in braneworld scenarios the
matter is to be confined to the four-dimensional universe), so
that all components of $T_{AB}$ are set to zero except the
spacetime components \cite{dar}. We assume the energy-momentum
tensor of spacetime to be an exotic fluid  $\chi$ with the
equation of state
\begin{equation}
p_\chi=\left(\frac{m}{3}-1\right)\rho_\chi,
\end{equation}
($p_\chi$ and $\rho_\chi$ are the pressure and the energy density
of the fluid, parameter $m$ has value between 0 and 2).

\subsection{Classical Model}

Dimensionally extended Einstein-Hilbert action
(without a higher-dimensional cosmological term) is
\begin{equation}
S=\int \sqrt{-g}\tilde Rdtd^3Rd^D\rho+S_m=\kappa\int dt L
\end{equation}
where $\kappa$ is an irrelevant constant, $\tilde R$ is the scalar
curvature of the metric, so we can read off the lagrangian of the
model (for flat internal space)
\begin{equation}
L=\frac{1}{2\tilde N}Ra^D\dot R^2 + \frac{D(D-1)} {12\tilde
N}R^3a^{D-2}\dot a^2 + \frac{D}{2\tilde N}R^2a^{D-1}\dot R\dot a -
\frac{1}{2}k\tilde N R a^D + \frac{1}{6}\tilde N\rho_\chi R^3a^D.
\end{equation}
For closed universe ($k=1$), substitution of the equation of state
in the continuity equation
\begin{equation}
\dot\rho_\chi R+3(p_\chi+\rho_\chi)\dot R=0
\end{equation}
leads to the energy density in form
\begin{equation}
\label{4.29}
\rho_\chi(R)=\rho_\chi(R_0)
\left(
\frac{R_0}{R}
\right)^m,
\end{equation}
where $R_0$ is the value of scaling factor in arbitrary reference
time. If we define cosmological constant as
$\Lambda=\rho_\chi(R)$, lagrangian becomes
\begin{equation}
\label{4.30} L=\frac{1}{2\tilde N}Ra^D\dot R^2 +
\frac{D(D-1)}{12\tilde N}R^3a^{D-2}\dot a^2 + \frac{D}{2\tilde
N}R^2a^{D-1}\dot R\dot a - \frac{1}{2}\tilde N R a^D +
\frac{1}{6}\tilde N\Lambda R^3a^D.
\end{equation}
Growth of the scaling factor $R$, according to (\ref{4.29}) leads to the
decrease
of the cosmological constant by the relation
\begin{equation}
\Lambda(R)=\Lambda(R_0)\left(\frac{R_0}{R}\right)^m.
\end{equation}
This decaying $\Lambda$ term may also explain the smallness of the
present value of the cosmological constant since, as the universe
evolves form its small to large size, the large initial value of
$\Lambda$ decays to small values. If we take $m=2$, initial
condition for cosmological constant and scaling factor
$\Lambda(R_0)R_0^2=3$, for laps function $\tilde
N(t)=R^3(t)a^D(t)N(t)$, the Lagrangian (\ref{4.30}) becomes
\begin{equation}
\label{4.32}
L=
\frac{1}{2N}\frac{\dot R^2}{R^2}
+
\frac{D(D-1)}{12N}\frac{\dot a^2}{a^2}
+
\frac{D}{2N}\frac{\dot R\dot a}{Ra}.
\end{equation}
It should be noted that there are no parameters $k$ and $\Lambda$
in the lagrangian. That means  that although they are not zero in
this model, it is equivalent to a flat universe with zero
cosmological term. In another words, there is not difference
between four dimensional universe which looks flat and not filled
with fluid and closed universe filled with exotic fluid.

The solutions of corresponding equations of motion are
\begin{equation}
R(t)=C_1 e^{\alpha t},
\end{equation}
\begin{equation}
a(t)=C_2 e^{\beta t},
\end{equation}
where the constants $C_1, C_2, \alpha$ and $\beta$ depend of
initial conditions. It is a reasonable assumption that the size of
all spatial dimensions is the same at $t=0$. It may be assumed
that this size would be the Planck size, i.e. $R(0)=a(0)=l_P$.
Above solutions, can be read in terms of Huble parameter $H=\dot
R/R$ \cite{dar}
\begin{equation}
R(t)=l_Pe^{Ht},
\end{equation}
\begin{equation}
a(t)=l_Pe^{-Ht}.
\end{equation}
Depending of the dimensionality of the internal space we have,
\begin{equation}
R(t)=l_Pe^{Ht},
\end{equation}
\begin{equation}
a_\pm(t)=l_Pe^{\frac{2Ht}{D}[-1\pm\sqrt{1-\frac{2}{3}(1-\frac{1}{D})}]^{-1
}},
\end{equation}
for $D=1$ and
\begin{equation}
R_\pm(t)=l_Pe^{\frac{D\beta
t}{2}[-1\pm\sqrt{1-\frac{2}{3}(1-\frac{1}{D})}]},
\end{equation}
\begin{equation}
a(t)=l_Pe^{\beta t},
\end{equation}
for $D>1$. The solution corresponding to $D=1$ predicts an
accelerating (de Sitter) universe and a contracting internal
space, with exactly the same rates. In the case $D>1$ analysis is
complicated, but results are similar.

\subsection{Quantum Model}

Quantum solutions are obtained from the Wheeler-DeWitt equation
\begin{equation}
\label{wdw} H\Psi(R,a)=0,
\end{equation}
where $H$ is the Hamiltonian and $\Psi$ is the wave function of
the universe. For this model above equation is read in the new
variables ($X=\ln R$ and $Y=\ln a$)
\begin{equation}
\left[ (D-1)\frac{\partial^2}{\partial X^2} +
\frac{6}{D}\frac{\partial^2}{\partial Y^2} -
6\frac{\partial}{\partial X}\frac{\partial}{\partial Y} \right]
\Psi(X,Y)=0.
\end{equation}
By the new change $x=X\frac{3}{D+3}+Y\frac{D}{D+3}$,
$y=\frac{X-Y}{D+3}$, Wheeler-DeWitt equation takes a simple form
\begin{equation}
\label{above} \left[ -3\frac{\partial^2}{\partial x^2} +
\frac{D+2}{D}\frac{\partial^2}{\partial y^2} \right] \Psi(x,y)=0.
\end{equation}
Equation (\ref{above}) has the four possible solutions \cite{dar}
\begin{equation}
\label{+-1}
\Psi_D^\pm(x,y)=
A^\pm
e^{\pm\sqrt{\frac{\gamma}{3}}x\pm\sqrt{\frac{\gamma D}{D+2}}y},
\end{equation}
\begin{equation}
\label{+-2}
\Psi_D^\pm(x,y)=
B^\pm
e^{\pm\sqrt{\frac{\gamma}{3}}x\mp\sqrt{\frac{\gamma D}{D+2}}y},
\end{equation}
where $A^\pm$ and $B^\pm$ are the normalization constants. It is
possible to impose the boundary conditions to get a
$\Psi_D(R,a)=0$. For the further details see \cite{dar}.

\section{(4+D)-Dimensional Model Over the Field of p-Adic Numbers}

Consideration of (4+D)-dimensional Kaluza-Klein model over the
field $Q_p$ will be started from the lagrangian in the form
(\ref{4.32}). All quantities in this Lagrangian will be treated as
the $p$-adic ones. Taking again replacement $X=\ln R$ and $Y=\ln
a$, it becomes
\begin{equation}
L= \frac{1}{2N}\dot X^2 + \frac{D(D-1)}{12N}\dot Y^2 +
\frac{D}{2N}\dot X\dot Y.
\end{equation}
The corresponding $p$-adic equations of motion are
\begin{equation}
\ddot{X}+\frac{D}{2}\ddot{Y}=0,\ \ \
\ddot{X}+\frac{D-1}{3}\ddot{Y}=0.
\end{equation}
It is not difficult to see that above system can be rewritten in
the following form
\begin{equation}
\label{jossamoova}
 \ddot{X}=0,\ \ \  \ddot{Y}=0.
\end{equation}
If we omit from consideration pseudoconstant solutions and
concentrate on the analytical ones we get $X(t)=C_1t+C_2$,
$Y(t)=C_3t+C_4$. The calculation of the $p$-adic classical action
gives
$$
\bar S_p(X'',Y'',N;X',Y',0)
$$
\begin{equation}
= \frac{1}{2N}(X''-X')^2 + \frac{D(D-1)}{12N}(Y''-Y')^2 +
\frac{D}{2N}(X''-X')(Y''-Y').
\end{equation}
Because this action is quadratic with respect to both variables
$X$ and $Y$, we can write down the kernel of $p$-adic operator of
evolution \cite{fil}
$$
{\cal K}_p(X'',Y'',N;X',Y',0)
$$
\begin{equation}
= \lambda_p \left( \frac{D(D+2)}{48N^2} \right) \left|
\frac{D(D+2)}{12N^2} \right|_p \chi_p(-\bar
S_p(X'',Y'',N;X',Y',0))
\end{equation}

Let use again the change $x=X\frac{3}{D+3}+Y\frac{D}{D+3}$,
$y=\frac{X-Y}{D+3}$ to separate variables and make further
analyzes of this model rather simple. In these variables the
classical action and the kernel of evolution operator read
$$
\bar S_p(x'',y'',N;x',y',0)
$$
\begin{equation}
=
\frac{1}{2N}(1+\frac{D(D+5)}{6})(x''-x')^2
-
\frac{1}{2N}D(D+3)(y''-y')^2,
\end{equation}
$$
{\cal K}_p(x'',y'',N;x',y',0)= \lambda_p \left( \frac{6+D(D+5)}{6}
\right) \lambda_p \left( - \frac{D(D+3)}{12N} \right)
$$
\begin{equation}
\label{kernel} \times \left| \frac{D(D+3)}{2N^2} \left(
1+\frac{D(D+5)}{6} \right) \right|_p^{1/2} \chi_p(-\bar
S_p(x'',y'',N;x',y',0)).
\end{equation}

Now, we can examine when $p$-adic wave function has the form
corresponding to the simplest ground state \cite{kon}
\begin{equation}
\label{pretposlednja}
\Psi_p(x,y)=\Omega(|x|_p)\Omega(|y|_p).
\end{equation}
Putting kernel of the operator of evolution (\ref{kernel}) in
equation (\ref{6.1}) we get that the required state exists if both
conditions
\begin{equation}
\label{5.52} |N|_p\leq \left|1+\frac{D(D+5)}{6}\right|_p, \ \ \
|N|_p\leq|D(D+3)|_p,\ p\neq2,
\end{equation}
are fulfilled. To answer the question: "can this conditions be
useful in determination of dimensionality of the internal space?"
needs further careful analysis.

Going back to the "old variables", $p$-adic ground state wave
function for our model is
\begin{equation}
\label{poslednja}
\Psi_p(x,y) = \Omega \left( \left| \left(
1-\frac{D}{D+3} \right)X + \frac{D}{D+3}Y \right|_p \right) \Omega
\left( \left| \frac{X-Y}{D-3} \right|_p \right).
\end{equation}
We can also write down the solutions in the variables $R$ and $a$.

\section{Conclusion}

In this paper, we demonstrate how a $p$-adic version of the
quantum (4+D)-Kaluza-Klein model with exotic fluid can be
constructed. It is an exactly soluble model. From equations
(\ref{+-1}), (\ref{+-2}) and (\ref{pretposlednja}), i.e.
(\ref{poslednja}), it is possible to construct an adelic model
too, i.e. model which unifies standard and all $p$-adic models
[12]. The investigation of its possible physical implication and
discreteness of space-time deserves much more attention and room.

Let us  note that adelic states for the (4+D)-dimensional
Kaluza-Klein cosmological model (for any D which satisfies
(\ref{5.52})) exist in the form
\begin{equation}
\label{7.1}
\Psi_S(x,y)=\Psi_{D,\infty}^\pm(x_\infty,y_\infty)
\prod_{p\in S} \Psi_p (x_p,y_p)
\prod_{p\notin S} \Omega(|x_p|_p)\Omega(|y_p|_p),
\end{equation}
\noindent where $\Psi_{D,\infty}^\pm(x_\infty,y_\infty)$ are the
corresponding real counterparts of the wave functions of the
universe. In the ground state wave functions $\Psi_p (x_p,y_p)$
are proportional to $\Omega(p^\nu |x_p|_p)$ $\times\Omega(p^\mu
|y_p|_p)$ or to $\delta (p^\nu -|x_p|)\delta (p^\mu -|y_p|)$.
Adopting the usual probability interpretation of the wave
function (\ref{7.1}), we have
\begin{equation}
\label{7.2}
\left|
\Psi_S(x,y)
\right|_\infty^2 =
\left| \Psi_{D,\infty}^\pm(x_\infty,y_\infty) \right|_\infty^2
\prod_{p\in S} |\Psi_p(x_p,y_p)|_\infty^2
\prod_{p\not\in S}
\Omega(|x_p|_p)\Omega(|y_p|_p),
\end{equation}
\noindent because $(\Omega(|x|_p))^2=\Omega(|x|_p)$.

As a consequence of $\Omega$-function properties, at the rational
points $x,y$ and in the (special) vacuum state ($S =\emptyset$,
i.e. all $\Psi_p(x_p,y_p)=\Omega(|x_p|_p)\Omega(|y_p|_p)$, we find
\begin{equation}
\label{7.3} \left| \Psi(x,y) \right|_\infty^2= \cases
{\left|\Psi_{D,\infty}^\pm(x,y)\right|_\infty^2, &$x,y\in Z$,\cr
0, &$x,y\in Q\backslash Z$.\cr}
\end{equation}
This result leads to some discretization of minisuperspace
coordinates $x,y$. Namely, probability to observe the universe
corresponding to our minisuperspace model is nonzero only in the
integer points of $x$ and $y$. Keeping in mind that
$\Omega$-function is invariant with respect to the Fourier
transform, this conclusion is also valid for the momentum space.
Note that this kind of discreteness depends on adelic quantum
state of the universe. When some $p$-adic states are different
from $\Omega(|x|_p)\Omega(|y|_p)$ ($S\neq \emptyset$), then the
above adelic discreteness becomes less transparent.

Performing the integration in (\ref{7.2}) over all $p$-adic
spaces, and having in mind that eigenfunctions should be normed to
unity, one recovers the standard effective model over real space.
However, if the region of integration is over only some parts of
$p$-adic spaces, then the adelic approach manifestly exhibits
$p$-adic quantum effects. Since the Planck length is here the
natural one, the adelic minisuperspace models refer to the Planck
scale.

Further investigations will be focused on determination of
conditions for existence of the ground states in the form
$\Omega(p^\nu|x|_p)\Omega(p^\mu|y|_p)$ and $p$-adic delta
function. We should emphasize  that investigation of
dimensionality $D$ of internal space from the conditions
(\ref{5.52}) and pseudoconstant solutions of the equation
(\ref{jossamoova}) deserves attention. It could additionally
contribute to better understanding of the model, especially from
its $p$-adic sector.

\section*{Acknowledgments}
The authors would like to thank the organizers of the "2nd
Conference on Theoretical Physics: Titeica-Markov Sympozium" for
the kind invitation to give this talk. Lj. Nesic is thankful to D.
Greku and M. Visinescu for their kind hospitality during the
Conference. The research of both authors was supported by the
Serbian Ministry of Science and Technology, Project No 1643.

\end{document}